\documentclass[a4paper,fleqn,usenatbib]{mnras}

\usepackage{newtxtext,newtxmath}

\usepackage[T1]{fontenc}
\usepackage{ae,aecompl}

\usepackage{graphicx}	
\usepackage{amsmath}	
\usepackage{amssymb}	
\usepackage{siunitx}    
\usepackage{bbm} 		
\usepackage{xcolor}		
\usepackage{soul}		
\usepackage{booktabs}	
\usepackage{array}		
\usepackage[inline]{enumitem}	
\usepackage{multirow}   
\usepackage{etoolbox}   

\newcommand{\xp}{X_{\textrm{p}}} 
\newcolumntype{L}[1]{>{\raggedright\let\newline\\\arraybackslash\hspace{0pt}}m{#1}}
\newcolumntype{C}[1]{>{\centering\let\newline\\\arraybackslash\hspace{0pt}}m{#1}}
\newcolumntype{R}[1]{>{\raggedleft\let\newline\\\arraybackslash\hspace{0pt}}m{#1}}

\makeatletter
\patchcmd{\NAT@citex}
  {\@citea\NAT@hyper@{%
     \NAT@nmfmt{\NAT@nm}%
     \hyper@natlinkbreak{\NAT@aysep\NAT@spacechar}{\@citeb\@extra@b@citeb}%
     \NAT@date}}
  {\@citea\NAT@nmfmt{\NAT@nm}%
   \NAT@aysep\NAT@spacechar\NAT@hyper@{\NAT@date}}{}{}
\patchcmd{\NAT@citex}
  {\@citea\NAT@hyper@{%
     \NAT@nmfmt{\NAT@nm}%
     \hyper@natlinkbreak{\NAT@spacechar\NAT@@open\if*#1*\else#1\NAT@spacechar\fi}%
       {\@citeb\@extra@b@citeb}%
     \NAT@date}}
  {\@citea\NAT@nmfmt{\NAT@nm}%
   \NAT@spacechar\NAT@@open\if*#1*\else#1\NAT@spacechar\fi\NAT@hyper@{\NAT@date}}
  {}{}
\makeatother

\title[Quasiperiodic glitches from a Poisson process]{Generating quasiperiodic pulsar glitches using a state-dependent Poisson process}

\author[Carlin et al.]{
J. B. Carlin,$^{1}$
A. Melatos$^{1,\,2}$
\\
$^{1}$School of Physics, University of Melbourne, Parkville, VIC 3010, Australia\\
$^{2}$Australian Research Council Centre of Excellence for Gravitational Wave Discovery (OzGrav), \\ \ University of Melbourne, Parkville, VIC 3010, Australia\\
}

\date{Accepted XXX. Received YYY; in original form ZZZ}

\pubyear{2018}

\begin{document}
\label{firstpage}
\pagerange{\pageref{firstpage}--\pageref{lastpage}}
\maketitle

\begin{abstract}
Glitching pulsars fall broadly into two statistical classes: those with Poisson-like waiting times and power-law sizes, and those with unimodal waiting times and sizes. Previous glitch modeling based on a state-dependent Poisson process readily generates Poisson-like behaviour but struggles to produce unimodal waiting times or sizes. Here it is shown that, when some of the inputs to the model are modified, both classes of statistical behaviour can be reproduced by varying a single control parameter related to the spin-down rate. The implications for past and future glitch observations and the underlying microphysical mechanism are explored briefly.
\end{abstract}

\begin{keywords}
pulsars: general -- stars: neutron -- stars: rotation -- methods: statistical
\end{keywords}



\section{Introduction}
\label{sec:intro}
Rotation-powered pulsars are observed to undergo impulsive spin-up events called ``glitches'' \citep{Melatos2008, Espinoza2011}, interrupting the monotonic spin down caused by electromagnetic braking \citep{Taylor1993}. While many statistical analysis of populations of glitching pulsars have been carried out ({see \citealp{Morley1993}; \citealp{Shemar1996}; \citealp{Lyne2000}, \citealp{Fuentes2017}; \citealp{Fuentes2018}; among others}), studies of the glitch statistics for individual pulsars have been stymied by small data sets; the five most prolific glitchers have between 20 and 42 glitches recorded up to 2018 May 28. Nevertheless, as these numbers grow, it is becoming possible to disaggregate the data and generate statistically meaningful probability density functions (PDFs) for the  glitch sizes and waiting times \citep{Ashton2017, Howitt2018}. Most pulsars with adequate sample sizes exhibit Poisson-like (i.e. exponentially distributed) waiting times  and scale-invariant (i.e. power-law distributed) event sizes. However, some pulsars show signs of quasiperiodicity in their glitch activity. Specifically, PSR J0835$-$4510 and PSR J0537$-$6910 have non-monotonic waiting time distributions \citep{Melatos2008, Espinoza2011, Howitt2018}.

Poisson-like and quasiperiodic behaviours are both consistent with a system existing in a state of self-organized criticality, in which stress accumulates under the action of a slow, global driver and is released sporadically by impulsive, stick-slip events \citep{Jensen1998, Melatos2008}. \citet{Bak1987} proposed a sandpile automaton as an idealized example of a self-organised critical system. Sandpiles exhibit power-law avalanche sizes and Poisson-like waiting times and fluctuate around a critical slope in a state of marginal stability. Experimental studies of real sandpiles often demonstrate large, system-spanning, quasiperiodic avalanches, and smaller avalanches in between \citep{Bretz1992, Rosendahl1993}. This distinction between large and small events also arises in seismology. Omori's Law describes the phenomenon of small earthquakes occurring at an increased rate after large earthquakes \citep{Omori1894, Utsu1995}. 

The exact physical mechanism that triggers glitches is unknown. Most microphysical models postulate that stress builds up in the system, as electromagnetic braking increases the elastic stress in the crust and/or crust-superfluid differential rotation. The stress is released spasmodically via superfluid vortex avalanches \citep{Anderson1975, Warszawski2011, Warszawski2012}, and/or starquakes \citep{Larson2002, Middleditch2006, Negi2011, Morley2018}, among other possibilities; see \citet{Haskell2015} for a modern review. To simulate and understand stress-release mechanisms of this kind, \citet{Fulgenzi2017} modelled long-term glitch activity as a mean-field, state-dependent Poisson process, generalising work done by \citet{Daly2006} and \citet{Wheatland2008}, who developed versions of the model in the context of forest fires and solar flares respectively. This class of meta-model is agnostic about the microphysics; it applies equally to superfluid vortex avalanches and starquakes, for example. It makes microphysics-independent, falsifiable predictions regarding glitch observables, e.g. correlations between sizes and waiting times \citep{Melatos2018}. 

Previous analyses of the state-dependent Poisson model have demonstrated that it generates power-law sizes and exponential waiting times naturally, when the spin-down rate is below a critical threshold \citep{Fulgenzi2017, Melatos2018}. Above the threshold, the model generates identical size and waiting time PDFs, as seen in quasiperiodic glitchers, but the functional form does not match observations; the model outputs power laws, whereas the data reveal Gaussian-like unimodal PDFs. It is therefore interesting to ask: can one modify the ingredients of the state-dependent Poisson model such that it reproduces both the observed Poisson-like and quasiperiodic behaviours by varying a single parameter --- the spin-down rate, normalized as discussed by \citet{Fulgenzi2017} --- below and above the threshold respectively? Answering this question systematically is the goal of this paper.

The paper is organized as follows. The state-dependent Poisson process model is described in detail in Section \ref{sec:sdpp}. In Section \ref{sec:eta} we test several forms of the conditional jump size distribution, a key input into the model, in an effort to generate quasiperiodic glitches. In Section \ref{sec:lambda} we test several forms of the Poisson rate function, another key model input. The observational implications are discussed briefly in Section \ref{sec:discussion}.

\section{State-dependent Poisson process}
\label{sec:sdpp}
\subsection{Equation of motion}
\citet{Fulgenzi2017} modelled long-term glitch activity as a mean-field, state-dependent Poisson process \citep{Cox1955, Daly2007, Wheatland2008}. In this model, the instantaneous glitch rate, $\lambda(t)$, is governed by a single variable $X(t)$. In the vortex avalanche model, $X(t)$ is the spatially averaged lag between the angular velocities of the rigid crust and the superfluid interior. In the crustquake model $X(t)$ measures the build-up of crustal stress or strain. We frame the presentation in terms of vortex avalanches, but emphasize that the meta-model transfers to any stick-slip stress-release process \citep{Melatos2018}. 

The angular velocity lag evolves according to
\begin{equation}
X(t) = X(0) + \frac{N_{\text{em}} t}{I_{\textrm{c}}} - \sum_{i=1}^{N(t)} \Delta X^{(i)} \ ,
\end{equation}
where $N_{\text{em}}$ is the electromagnetic spin-down torque acting on the crust, $I_{\textrm{c}}$ is the moment of inertia of the crust, $X(0)$ is an arbitrary initial condition, $N(t)$ is the number of glitches that have occurred up to time $t$, and $\Delta X^{(i)}$ is the size of the $i$-th glitch. By conservation of angular momentum we can relate $\Delta X^{(i)}$ to $\Delta \nu^{(i)}$, the (observable) spin frequency gained by the crust at each glitch, viz.
\begin{equation}
\label{eq:delx_ratio}
\Delta X^{(i)} = - \frac{2\pi (I_{\textrm{c}} + I_{\textrm{s}}) \Delta \nu^{(i)}}{I_{\textrm{s}}} \ ,
\end{equation}
where $I_{\textrm{s}}$ is the moment of inertia of the superfluid interior.

Both $N(t)$ and $\Delta X^{(i)}$ are random variables. The avalanche sizes $\Delta X^{(i)}$, while random, depend on $X(t)$ before the glitch. $N(t)$ is a Poisson counting process governed by the waiting time distribution discussed in Section \ref{sec:times}.

\subsection{Sizes}
\label{sec:sizes}
The size of a glitch is governed statistically by $\eta(\Delta X \mid \xp)$, a conditional jump distribution that depends on the lag immediately before the glitch, $\xp$. We posit that glitches always decrease the lag but cannot make it negative. The function $\eta(\Delta X \mid \xp)$ is one of the ingredients we seek to modify in this paper in an effort to generate quasiperiodic behaviour in a natural way.

One choice for $\eta$, suggested by Gross-Pitaevskii simulations\footnote{The Gross-Pitaevskii equation describes the evolution of a zero temperature Bose-Einstein condensate, which is often used as an idealized model of a neutron star superfluid. See the recent review by \citet{Haskell2015} for details.} \citep{Warszawski2011, Warszawski2013} and by analogy with other systems that exhibit signs of self-organised criticality \citep{Jensen1998, Aschwanden2018}, is a power law of the form
\begin{equation}
\label{eq:eta}
 \eta(\Delta X \mid \xp) = \left[\int_{0}^{\xp}d\xi\, g(\xi, \xp) \right]^{-1} g(\Delta X, \xp) \ ,
\end{equation}
where we define for convenience
\begin{equation}
\label{eq:gpow}
g(\xi, \xp) = \xi^{-1.5} H(\xi - \beta \xp) H(\xp - \xi)\ .
\end{equation}
Equation \eqref{eq:eta} is automatically normalised in the domain $0 \leq \Delta X \leq \xp$ by the pre-factor. The exponent $-1.5$ in $g(\xi, \xp)$ is typical of avalanche processes and consistent with Gross-Pitaevskii simulations \citep{Jensen1998}. The factor $\beta < 1$ in the Heaviside function $H$ ensures that the integral converges by setting a minimum glitch size as a fraction of $\xp$. It is obligatory for the power law in \eqref{eq:gpow} but not for every other functional form. The impact of the functional form on the long-term statistics is explored in Section \ref{sec:eta}. There is freedom to choose $\eta(\Delta X \mid \xp)$, as there is currently no possibility of observing it directly in a neutron star.

\subsection{Waiting times}
\label{sec:times}
The instantaneous glitch rate is modelled as a time-dependent Poisson process, whose rate is a function of the instantaneous lag, i.e. $\lambda[X(t)]$. We assume the rate is a monotonically increasing function of $X(t)$ and diverges at a critical lag, $X_{\textrm{cr}}$, where a glitch is certain to occur. The function $\lambda(X)$, like $\eta(\Delta X \mid \xp)$, is one of the ingredients we seek to modify in this paper to reproduce quasiperiodic glitch activity. 

As $\lambda\left[X(t)\right]$ evolves deterministically between glitches, we can use the standard PDF for a variable-rate Poisson process to pick the waiting times, $\Delta t$, between glitches, conditional on the deterministic evolution of the lag in the system, $X(t)$, after the previous glitch:
\begin{equation}
\label{eq:times}
\textit{p}\left[\Delta t \mid X(t)\right] = \lambda\left[X(t) + \frac{N_{\text{em}}\Delta t}{I_{\textrm{c}}}\right]\exp\left\{-\int_{t}^{t + \Delta t} dt' \lambda\left[X(t')\right]\right\} \ .
\end{equation}

For the analysis in Section \ref{sec:eta} we follow \citet{Fulgenzi2017} in choosing the following phenomenological functional form:
\begin{equation}
\label{eq:rate}
\lambda(X) = \lambda_0 \left(1 - X / X_{\textrm{cr}} \right)^{-1} \ .
\end{equation}
Here $\lambda_0$ is a reference rate, e.g. $\lambda_0 = \lambda(X_{\textrm{cr}}/2)/2$, and $X_{\textrm{cr}}$ is a complex function of the nuclear physics of vortex pinning. \citet{Fulgenzi2017} argued that the exact functional form of \eqref{eq:rate} does not affect the long-term statistics, as long as $\lambda(X)$ diverges at $X=X_{\textrm{cr}}$ and increases monotonically with $X$. We test this claim further in Section \ref{sec:lambda} by investigating the interaction between the rate function, the choice of conditional jump size distribution, and other control parameters.

\subsection{Dimensionless variables}
\label{sec:dimen}
The model is expressed usefully in dimensionless variables by setting $\tilde{X} = X / X_{\textrm{cr}}$ and $\tilde{t} = N_{\text{em}}t/(X_{\textrm{cr}} I_{\textrm{c}})$. The main control parameter is $\alpha$, which is introduced through the dimensionless version of \eqref{eq:rate},
\begin{equation}
\label{eq:ndim_rate}	
\lambda(\tilde{X}) = \frac{\alpha}{1 - \tilde{X}}\ ,
\end{equation}
with
\begin{equation}
\alpha = \frac{I_{\textrm{c}} X_{\textrm{cr}} \lambda_0}{N_{\text{em}}} \ .
\end{equation}
\citet{Fulgenzi2017} showed that the model output is classified into two regimes: $\alpha \gtrsim \alpha_c(\beta) \approx \beta^{-1/2}$, called the slow spin-down regime, which generates exponentially distributed waiting times and power-law  sizes; and $\alpha \lesssim \alpha_c(\beta)$, where the sizes and waiting times are identically distributed as power laws; see Section 4 in \citet{Fulgenzi2017}.

\subsection{Monte Carlo simulations}
\label{sec:autom}
The evolution of $X$ can be modelled through a simple five-step Monte Carlo automaton:\footnote{Tildes are dropped here and henceforth in this paper for clarity.}
\begin{itemize}[labelindent=0pt,labelwidth=1em,leftmargin=!]
	\item[1.] Pick a random $\Delta t$ from \eqref{eq:times} given the current lag $X$.
	\item[2.] Update the lag to $X + \Delta t$ to account for the deterministic evolution up to the glitch.
	\item[3.] Pick a random $\Delta X$ from \eqref{eq:eta} given the lag just prior to the glitch.
	\item[4.] Subtract $\Delta X$ from the lag.
	\item[5.] Repeat from step 1.
\end{itemize}
Random numbers are picked using a rejection method. The method is useful when the PDF has a finite upper bound. It handles functional forms in \eqref{eq:eta} and \eqref{eq:times} that are not easily integrable or invertible, as required by the standard inverse cumulative algorithm \citep{Press2007}.

\section{Jump size distribution $\eta(\Delta X \mid \xp)$}
\label{sec:eta}
The meta-model accommodates any physically plausible distribution for the jump sizes; the power law in \eqref{eq:gpow} is not the only possible functional form. In the following analysis we keep the following two fundamental restrictions: the lag is never negative and it always decreases at a glitch, i.e. 
\begin{enumerate*}
\item $\eta(\Delta X \mid \xp) = 0 $ for $\Delta X > \xp$ and
\item  $\eta(\Delta X \mid \xp) = 0$ for $\Delta X \leq 0$.
\end{enumerate*}

\subsection{Functional form}
With the above restrictions in mind, we posit a suite of alternative jump size distributions, broadly categorized by two traits: ``monotonic'' or ``unimodal'', and ``fixed'' or ``stretchable'' (for those that are not scale-invariant). The categorization attempts to provide a rough census of the function space covered by $\eta(\Delta X \mid \xp)$. For completeness we also test some ``quirky'' functions, e.g. trigonometric and horseshoe-shaped. The functional forms of all jump size distributions tested in this paper are tabulated in Table \ref{tab:eta}.

\begin{figure}
	\centering
	\includegraphics[width=0.9\linewidth]{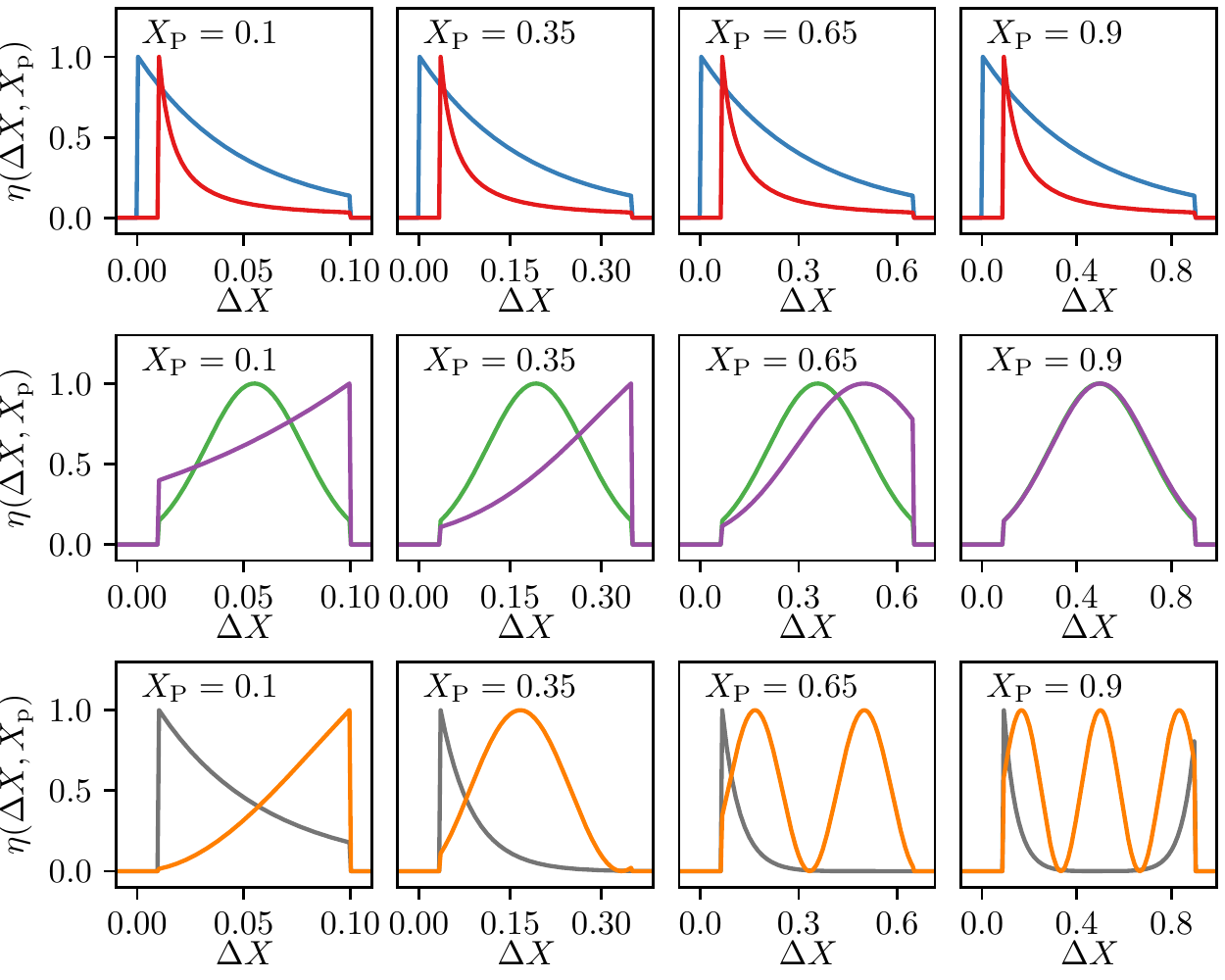}
	\caption{Examples of conditional jump distributions $\eta(\Delta X \mid \xp)$ distributions for $\xp$ (pre-glitch lag) increasing from 0.1 (left panels) to 0.9 (right panels). (Top row.) Power law (red curves) and stretchable exponential (blue). (Middle row.) Stretchable (green) and fixed (purple) Gaussian. (Bottom row.) Fixed trigonometric (orange) and fixed horseshoe (grey). The maximum of each curve has been arbitrarily scaled to unity to facilitate shape comparison.}
	\label{fig:g_4panel}
\end{figure}

A monotonic distribution is one that decreases with $\Delta X$. In contrast, a unimodal distribution is one with a single, well-defined peak, like a Gaussian. Two monotonic distributions are shown in the top row of panels in Figure \ref{fig:g_4panel}: the red curve is the power law defined in \eqref{eq:gpow}; the blue curve is a stretchable exponential. Both curves ``stretch'' with $\xp$, i.e. their shape is the same, regardless of $\xp$.

We classify a distribution as ``fixed'' if the shape does not shift with $\xp$, i.e. if we have
\begin{equation}
{g(\xi, \xp) = f(\xi) H(\xp - \xi)}\ ,
\end{equation}
for some arbitrary function $f$, where the Heaviside function ensures restriction (i) above is maintained. A ``stretchable'' distribution on the other hand does shift with $\xp$, such that its shape is maintained regardless of $\xp$. This is seen clearly in the middle row of panels in Figure \ref{fig:g_4panel}, where the green curves are stretchable, while the purple curves are fixed.

The bottom row of panels Figure \ref{fig:g_4panel} shows two ``quirky'' functional forms: a fixed trigonometric function (orange curves), and a fixed horseshoe-shaped function (grey curves).

Many of the functional forms require additional parameters to specify their shape as specified in Table \ref{tab:eta}, e.g. $\mu_\textrm{G}$ and $\sigma_\textrm{G}$ for the Gaussian, and $\beta_\textrm{E}$ for the exponential. In general, the specific choice of these parameters does not affect the overall shape of the long-term statistics produced by the model, i.e. the observable PDFs of waiting times, $\textit{p}(\Delta t)$, and sizes, $\textit{p}(\Delta X)$. 

\subsection{Monotonic versus unimodal}
\label{sec:unimodal}
\begin{figure}
	\centering
	\includegraphics[width=0.95\linewidth]{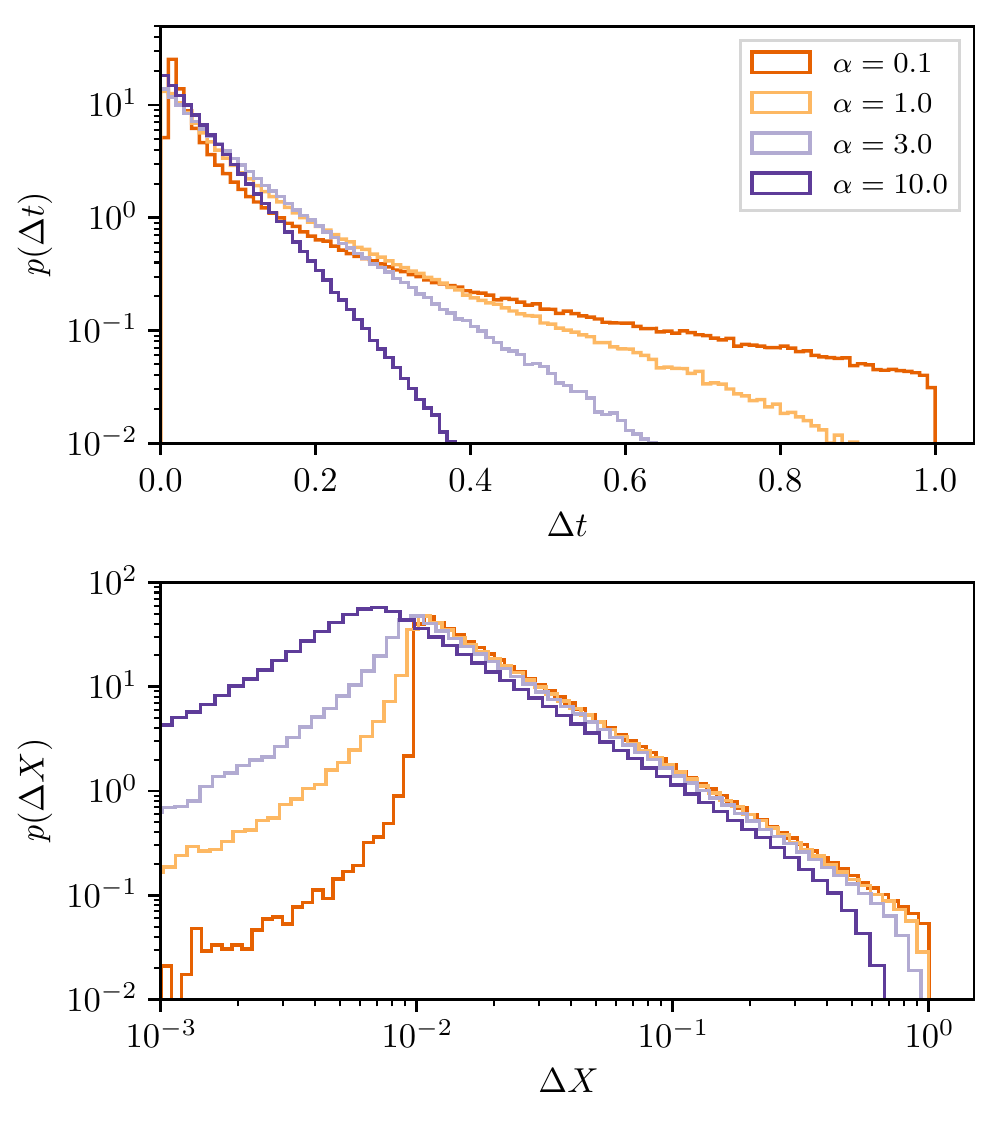}
	\caption{Long-term glitch statistics generated by the power-law jump size distribution defined in \eqref{eq:gpow}. (Top panel.) Waiting time PDF on log-linear axes. (Bottom panel.) Size PDF on log-log axes. All PDFs are generated using $N = 10^6$ glitches. Color legend for the control parameter $\alpha$ is shared between panels. Parameter: $\beta = 10^{-2}.$}
	\label{fig:std_pow}
\end{figure}

\begin{figure}
	\centering
	\includegraphics[width=0.95\linewidth]{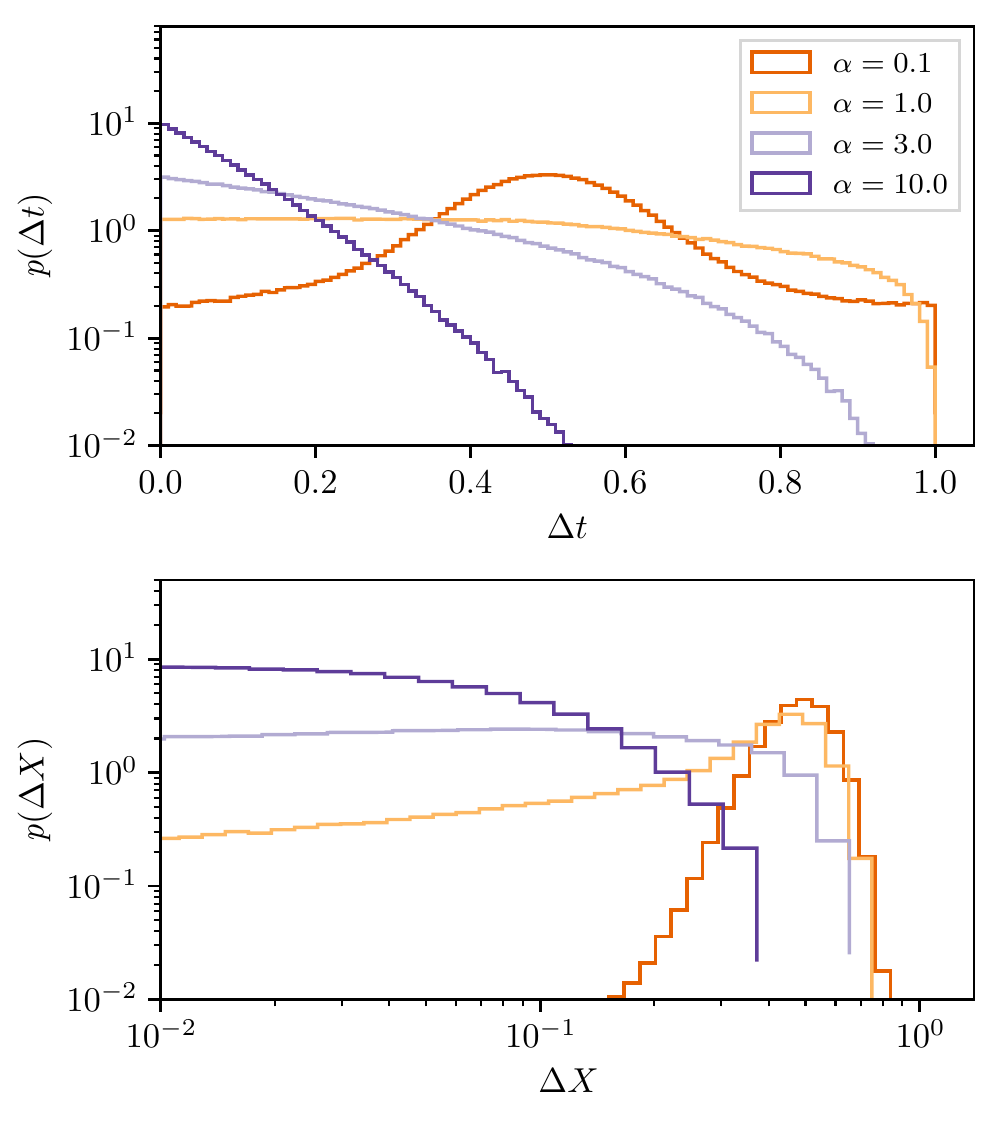}
	\caption{Long-term glitch statistics generated by the Gaussian jump size distribution defined in \eqref{eq:gausseta}. (Top panel.) Waiting time PDF on log-linear axes. (Bottom panel.) Size PDF on log-log axes. All PDFs are generated using $N = 10^6$ glitches. Color legend for the control parameter $\alpha$ is shared between panels. Parameters: $\mu_\textrm{G}=0.5$, $\sigma_\textrm{G}=0.125$.}
	\label{fig:std_gauss_mid_fixed}
\end{figure}

\begin{figure}
	\centering
	\includegraphics[width=0.95\linewidth]{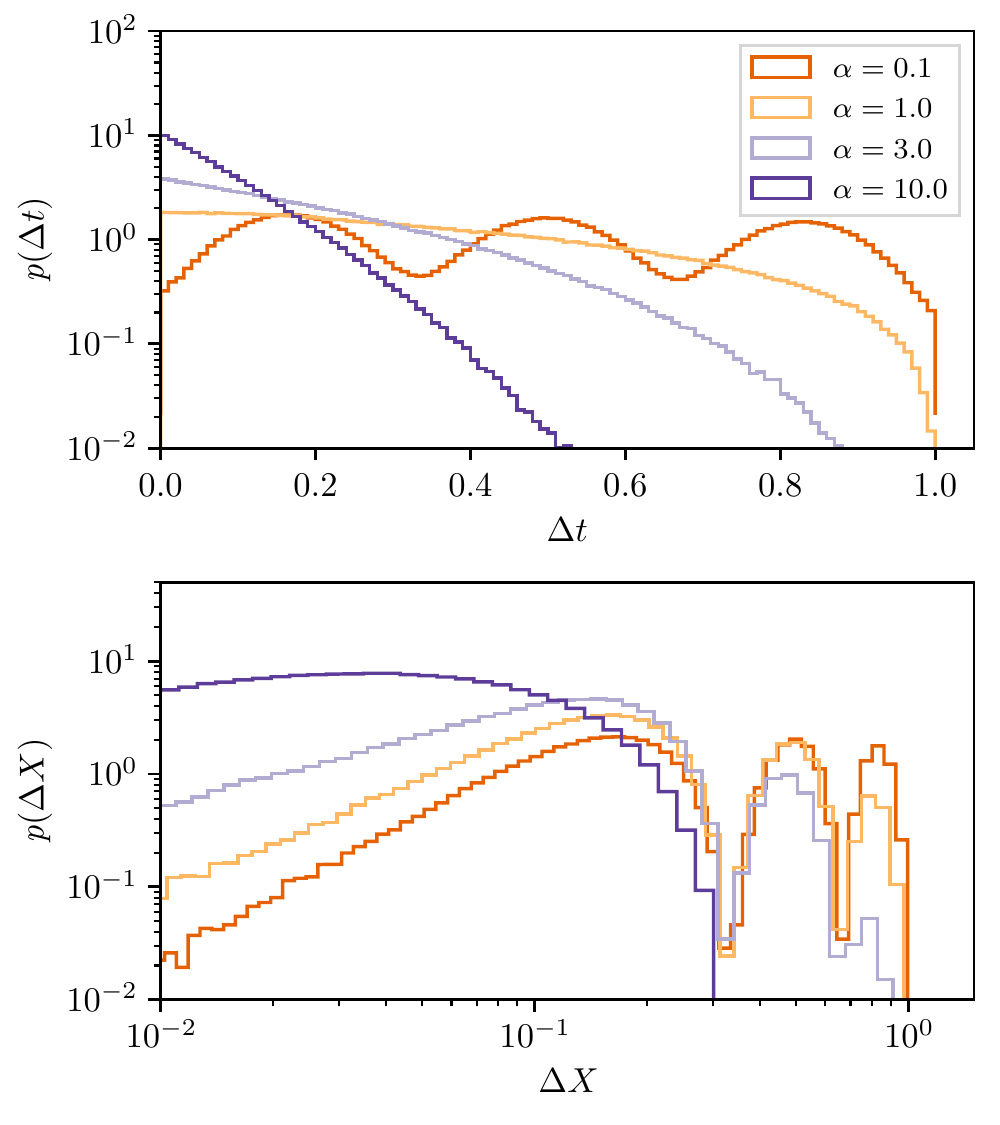}
	\caption{Long-term glitch statistics generated by the trigonometric jump size distribution defined in \eqref{eq:sin2eta}. (Top panel.) Waiting time PDF on log-linear axes. (Bottom panel.) Size PDF on log-log axes. All PDFs are generated using $N = 10^6$ glitches. Color legend for the control parameter $\alpha$ is shared between panels. Functional form: see Table \ref{tab:eta}.}
	\label{fig:std_sin2_fixed}
\end{figure}

The effect of $\eta(\Delta X, \xp)$ on $\textit{p}(\Delta t)$ and  $\textit{p}(\Delta X)$ is explored in Figures \ref{fig:std_pow}, \ref{fig:std_gauss_mid_fixed}, and \ref{fig:std_sin2_fixed}. We emphasise that $\textit{p}(\Delta t)$ and $\textit{p}(\Delta X)$ are not the same as equations \eqref{eq:times} and \eqref{eq:eta} respectively, as the latter are instantaneous PDFs, while the former are generated by Monte Carlo simulations, during which $X(t)$ and $\xp$ fluctuate stochastically. Figures \ref{fig:std_pow}--\ref{fig:std_sin2_fixed} are all constructed using the rate function \eqref{eq:ndim_rate}. The effect of modifying $\lambda(X)$ is studied in Section \ref{sec:lambda}.

Figure \ref{fig:std_pow} reproduces the results of \citet{Fulgenzi2017}, when $\eta(\Delta X, \xp)$ is a power law with exponent $-1.5$ and $\beta=10^{-2}$, as in \eqref{eq:gpow}. Exponential waiting times are seen for $\alpha \geq 3$, while  $\textit{p}(\Delta t)$ is a power law for $\alpha \leq 1$. This supports the conclusion of \citet{Fulgenzi2017}: there exist two regimes of activity, $\alpha \gtrsim \beta^{-1/2}$ (slow spin-down) and $\alpha \lesssim \beta^{-1/2}$ (fast spin-down). The size distribution is a power law over roughly two decades for all values of $\alpha$ tested. The lower cut-off is dictated by $\alpha$ and $\beta$. When $\alpha$ is low, the lag climbs higher on average before a glitch, magnifying the impact of the minimum glitch size, which is specified as a fraction of the lag prior to the glitch.

Figure \ref{fig:std_gauss_mid_fixed} shows the long-term statistics generated by a unimodal jump size distribution of the form
\begin{equation}
\label{eq:gausseta}
{\eta(\Delta X \mid \xp) \propto \exp\left[\frac{-(\Delta X - \mu_\textrm{G})^2}{\sigma_\textrm{G}^2}\right]H(\xp - \Delta X) H(\Delta X)}\ ,
\end{equation}
where the proportionality constant is set by normalization. This truncated Gaussian is drawn as the purple curves in the middle row of panels in Figure \ref{fig:g_4panel}. The dimensionless mean, $\mu_\textrm{G}$, and scale, $\sigma_\textrm{G}$, of the Gaussian are fixed at $0.5$ and $0.125$ respectively. Exponential waiting times are produced for $\alpha = 10$. The size PDF is not a power law for any $\alpha$. However it is monotonically decreasing and roughly scale invariant for $\alpha \gtrsim 3$. 

As $\alpha$ decreases, $\textit{p}(\Delta t)$ becomes more uniform and ultimately non-monotonic for $\alpha \lesssim 1$. In the fast spin-down (low $\alpha$) regime its shape resembles $\eta(\Delta X \mid \xp)$ in the limit $\xp \rightarrow 1$, because we have $\xp \approx 1$ just before every glitch. The waiting time roughly equals $\Delta X$ at the previous glitch, as the system recovers back to $X \approx 1$. The same behaviour is seen in the size PDF, which approaches a Gaussian as $\alpha$ decreases. 

Broadly speaking, adjusting $\mu_\textrm{G}$ and $\sigma_\textrm{G}$ does not change the shape of the results. Instead it shifts the mean and variance of the resultant PDFs. In the low $\alpha$ regime we find $\langle \Delta t \rangle$, $\langle \Delta X \rangle \rightarrow 1$ for $\mu_\textrm{G} \rightarrow 1$. Increasing $\sigma_\textrm{G}$ such that $\eta(\Delta X \mid \xp)$ is a broader distribution generates broader $\textit{p}(\Delta t)$ and $\textit{p}(\Delta X)$, as expected.

Figure \ref{fig:std_sin2_fixed} shows roughly similar results, when an oddly-shaped jump size distribution is chosen, e.g.
\begin{equation}
\label{eq:sin2eta}
{\eta(\Delta X \mid \xp) \propto \sin^2(3\pi \Delta X) H(\xp - \Delta X) H(\Delta X)}\ .
\end{equation}
There is no physical motivation behind this choice. Its purpose is to see how the model responds when pushed in nontraditional directions. In the fast spin-down regime the shape of $\eta(\Delta X \mid \xp)$ is reflected in both the waiting time and size PDF. In the slow spin-down regime exponential waiting times are seen, while the size PDF decreases monotonically, with a sharp upper cut-off at $\Delta X \approx \num{2e-1}$.

\subsection{Fixed versus stretchable}
\begin{figure}
	\centering
	\includegraphics[width=0.7\linewidth]{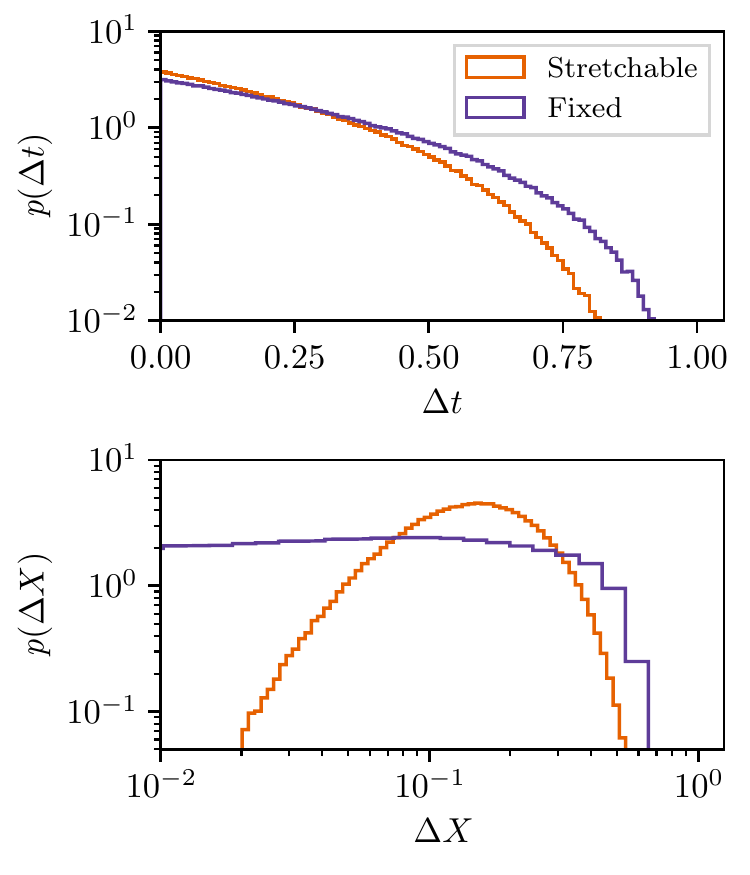}
	\caption{Long-term glitch statistics generated by the fixed Gaussian jump size distribution defined in \eqref{eq:gausseta} (purple curve), and the stretchable Gaussian jump size distribution defined in \eqref{eq:etagauss_stretch} (orange curve). (Top panel.) Waiting time PDF on log-linear axes. (Bottom panel.) Size PDF on log-log axes. All PDFs are generated using $N = 10^6$ glitches. Parameters: $\alpha = 3$, $\mu_\textrm{G} = 0.5$, $\sigma_\textrm{G}=0.125$.}
	\label{fig:stretch_fix_comp}
\end{figure}
Both the Gaussian and trigonometric jump size distributions discussed in Section \ref{sec:unimodal} are ``fixed'', in the sense that the parameters that describe their shape do not scale with $\xp$. What happens if the parameters are stretchable, meaning that the shape of the distribution stretches with $\xp$, e.g. the blue and green curves in Figure \ref{fig:g_4panel}? Figure \ref{fig:stretch_fix_comp} compares $\textit{p}(\Delta t)$ and $\textit{p}(\Delta X)$ at $\alpha=3$ for the Gaussian defined in \eqref{eq:gausseta} and the stretchable alternative
\begin{equation}
\label{eq:etagauss_stretch}
{\eta(\Delta X \mid \xp) \propto \exp\left[\frac{-(\Delta X - \mu_\textrm{G}\xp)^2}{(\sigma_\textrm{G}\xp)^2}\right]H(\xp - \Delta X) H(\Delta X)}\ .
\end{equation}
The parameters $\mu_\textrm{G}$ and $\sigma_\textrm{G}$ are set to $0.5$ and $0.125$ respectively for both \eqref{eq:gausseta} and \eqref{eq:etagauss_stretch}. The waiting time PDFs are similar for the two alternatives. The size PDF for the stretchable distribution is much narrower than for the fixed distribution. Most of the probability mass in the stretchable distribution lies halfway between 0 and $\xp$ making small glitches unlikely. The different behaviour between fixed and stretchable jump size distributions is broadly replicated for small and large $\alpha$ and various functional forms of $\eta(\Delta X \mid \xp)$.

\begin{table*}
\centering
\renewcommand{\arraystretch}{1.5}%
\begin{tabular}{L{8em} L{6.5em} l l L{6.7em} L{6.7em} L{6.7em}}
	\toprule
	Shape & Fixed (F) or stretchable (S) & $\eta(\Delta X \mid \xp)$ & $\textit{p}(\Delta t)$, $\alpha=10$ &  $\textit{p}(\Delta X)$, $\alpha=10$ & $\textit{p}(\Delta t)$, $\alpha=0.1$ &  $\textit{p}(\Delta X)$, $\alpha=0.1$\\
	\midrule
	Power law & N/A & $\Delta X ^{-1.5} H(\Delta X - \beta \xp)$ & Exponential & Power law & Power law & Power law\\
	Uniform & N/A & $1$ & Exponential & Monotonic decreasing & Uniform & Uniform\\
	Increasing power law & N/A & $\Delta X ^{1.5}$ & Exponential & Monotonic decreasing & Monotonic increasing & Monotonic increasing\\
	\multirow{2}{*}{Exponential} & S & $\exp\left(\frac{-\beta_\textrm{E}\Delta X}{\xp}\right)$ & \multirow{2}{*}{Exponential} & \multirow{2}{6.7em}{Monotonic decreasing} & \multirow{2}{*}{Exponential} & \multirow{2}{6.7em}{Monotonic decreasing} \\
	 & F & $\exp\left(-\beta_\textrm{E}\Delta X\right)$ &  &  &  &  \\
	\multirow{2}{*}{Gaussian} & S & $\exp{\left[\frac{-(\Delta X - \mu_\textrm{G} \xp)^2}{(\sigma_\textrm{G} \xp)^2}\right]}$ & \multirow{2}{*}{Exponential} & \multirow{2}{6.7em}{Monotonic decreasing} & \multirow{2}{*}{Unimodal} & \multirow{2}{*}{Unimodal}\\
	 & F & $\exp{\left[\frac{-(\Delta X - \mu_\textrm{G})^2}{(\sigma_\textrm{G})^2}\right]}$ &  &  &  & \\
	Trigonometric & F & $\sin^2(3\pi \Delta X)$ & Exponential & Monotonic decreasing & Non-monotonic & Unimodal \\
	Narrow horseshoe & F & $\left\{ \exp\left[\frac{(\Delta X - \mu_\textrm{H})^2}{(\sigma_\textrm{H})^2} \right] - 1 \right\} ^2$ & Exponential & Non-monotonic & Non-monotonic & Non-monotonic\\
	\bottomrule
\end{tabular}
\caption{Role of the conditional jump size distribution: functional forms tested and output PDFs (sizes and waiting times). The product of Heaviside functions, $H(\Delta X)H(\xp - \Delta X)$, that imposes the restrictions outlined in Section \ref{sec:sdpp} multiplies all the entries in the third column, along with a proportionality constant determined by normalization. The general trends noted in the final four columns are discussed in Sections \ref{sec:eta} and \ref{sec:discussion}.}
\label{tab:eta}
\end{table*}

\subsection{Summary}
Based on Figures \ref{fig:std_pow}--\ref{fig:stretch_fix_comp} and additional simulations not plotted here, the general behaviour can be broadly categorized against the criteria in Table \ref{tab:eta}. 
\begin{enumerate*}
\item In the high-$\alpha$ regime, all of the tested jump size distributions produce exponential waiting times and all have monotonically decreasing glitch size distributions, except for the ``quirky'' narrow horseshoe.
\item At high $\alpha$, none of the tested jump size distributions produce power-law distributed glitch sizes, except for \eqref{eq:gpow}. This includes monotonic $\eta$, such as the stretchable exponential (the blue curves in Figure \ref{fig:g_4panel}) which qualitatively resembles a power law.
\item At low $\alpha$, we find $\textit{p}(\Delta t) \approx \textit{p}(\Delta X) \approx \eta(\Delta X \mid \xp \rightarrow 1)$ for all functional forms.
\end{enumerate*}

\section{Rate function $\lambda(X)$}
\label{sec:lambda}
In an effort to produce quasiperiodic glitch activity, or a unimodal waiting time PDF, one may contemplate modifying the rate function defined in \eqref{eq:ndim_rate} instead of $\eta(\Delta X \mid \xp)$. \citet{Fulgenzi2017} found that a rate function of the form $\lambda[X(t)] = \alpha \tan\left[\pi X(t) / 2\right]$, which also diverges as $X(t) \rightarrow 1$, does not qualitatively change the long term statistics, when $\eta(\Delta X \mid \xp)$ is a power law. However, if the divergence at $X \rightarrow 1$ is steeper than in \eqref{eq:ndim_rate}, it is reasonable to expect some quasiperiodicity in the long-term statistics, as the system drives itself back to $X \approx 1$ faster. 

\begin{figure}
	\centering
	\includegraphics[width=0.95\linewidth]{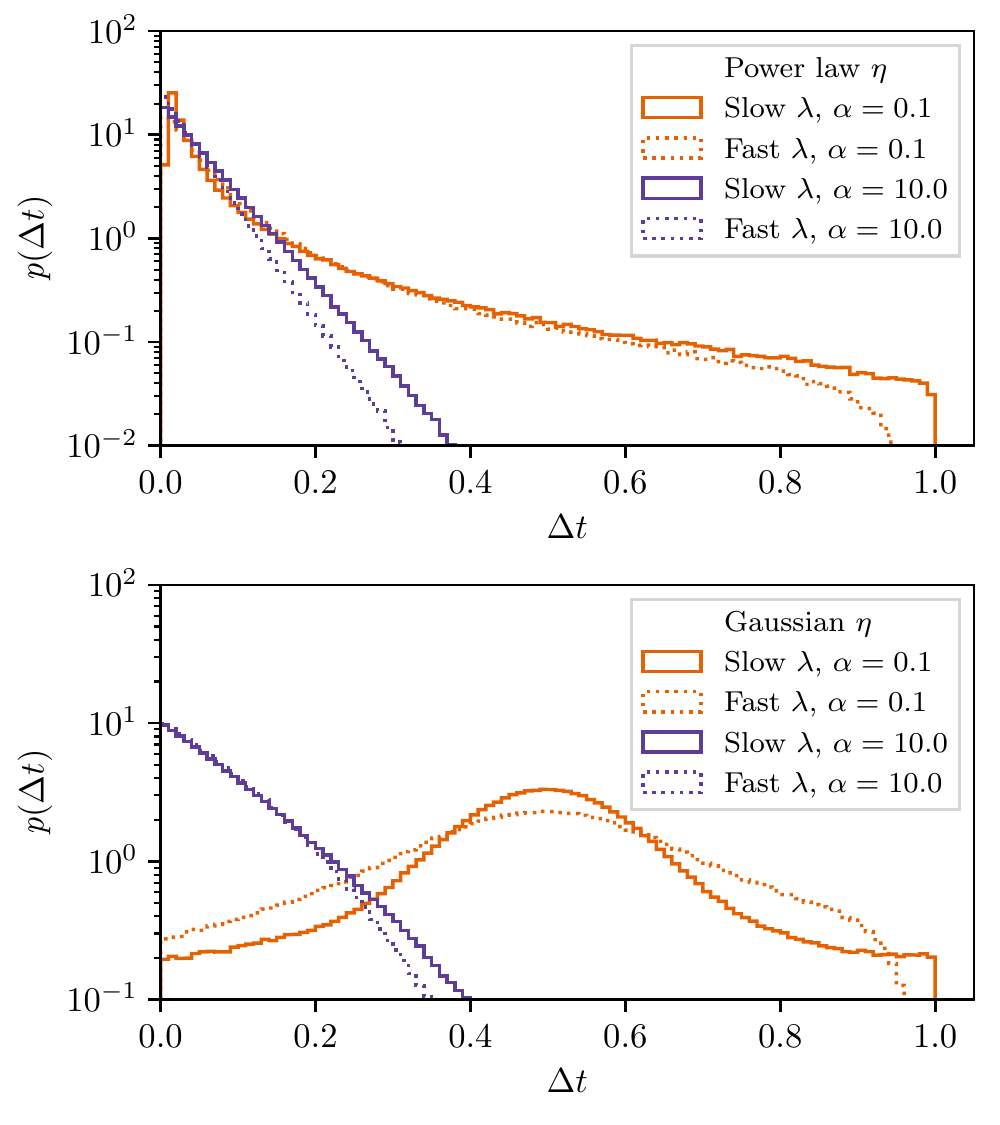}
	\caption{Waiting time PDFs on log-linear axes comparing the effect of the ``slow'' rate function defined in \eqref{eq:ndim_rate} (solid curves) and ``fast'' rate function defined in \eqref{eq:fastrate} (dashed curves). Orange curves correspond to low $\alpha$ and purple curves correspond to high $\alpha$. (Top panel.) Power law $\eta(\Delta X \mid \xp)$ defined in \eqref{eq:gpow} with $\beta = 10^{-2}$. (Bottom panel.) Gaussian $\eta(\Delta X \mid \xp)$ defined in \eqref{eq:gausseta} with $\mu_\textrm{G}=0.5$ and $\sigma_\textrm{G}=0.125$. All PDFs are generated using $N=10^6$ glitches.}
	\label{fig:lamcomp}
\end{figure}

We test the above hypothesis by proposing a ``fast'' alternative to \eqref{eq:ndim_rate}, i.e.
\begin{equation}
\label{eq:fastrate}
{\lambda\left[X(t)\right] = \frac{\alpha}{[1 - X(t)]^2}}\ .
\end{equation}
The exponent of $-2$ is arbitrary; the results do not change much if the divergence is steeper (i.e. higher exponent). Figure \ref{fig:lamcomp} shows that, to leading order, there is no change to the waiting time PDF when \eqref{eq:ndim_rate} is replaced by \eqref{eq:fastrate}. The result holds for $\alpha=0.1$, $\alpha=10$, Gaussian $\eta$, and power-law $\eta$. The top panel displays $\textit{p}(\Delta t)$ for the power-law $\eta$ defined in \eqref{eq:gpow}. The bottom panel displays $\textit{p}(\Delta t)$ for the Gaussian $\eta$ defined in \eqref{eq:gausseta}. In the top panel, \eqref{eq:fastrate} generates a thinner tail than \eqref{eq:ndim_rate}, but the general shape is the same. In the bottom panel, \eqref{eq:fastrate} produces more dispersion in $\Delta t$  than \eqref{eq:ndim_rate} at low $\alpha$ and again has a thinner tail at high $\alpha$.

The rate function, $\lambda[X(t)]$, is an input into the conditional waiting time PDF, $\textit{p}(\Delta t \mid X)$, defined in \eqref{eq:times}. For all $X$, $\textit{p}(\Delta t \mid X)$ is a monotonically increasing function of $\Delta t$ for $\alpha<1$ and a monotonically decreasing function of $\Delta t$ for $\alpha >1$. The changeover shifts to $\alpha = 2$ for the faster rate function \eqref{eq:fastrate}, independent of $\eta(\Delta X \mid \xp)$.

It may be tempting to reverse-engineer quasiperiodic waiting times by choosing a specific rate function that makes $\textit{p}(\Delta t \mid X)$ non-monotonic. One simple way to do this is to relax the requirement that the rate diverges as $X \rightarrow 1$, e.g.
\begin{equation}
\label{eq:xp1}
{\lambda[X(t)] = \alpha X(t)}\ .
\end{equation}
Equation \eqref{eq:xp1} leads to a skewed Gaussian for $\textit{p}(\Delta t \mid X)$ and an exponential for $\textit{p}(\Delta t)$ for many choices of $\alpha$ and $\eta(\Delta X \mid \xp)$. This means that \eqref{eq:xp1} does not naturally lead to two distinct classes of glitchers: ones with exponential waiting times and power law sizes, and those with unimodal waiting times and sizes. This is not to say that reverse-engineering quasiperiodicity using the rate function is impossible. A rate function that is sharply peaked at a certain lag, e.g. $\lambda[X(t)] = \delta [1 - X(t)]$, where $\delta$ is the Dirac delta function,  always triggers glitches at $X(t)=1$. Hence $\textit{p}(\Delta t)$ and $\textit{p}(\Delta X)$ are the same as $\eta(\Delta X \mid \xp \rightarrow 1)$, i.e. they are unimodal if $\eta(\Delta X \mid \xp \rightarrow 1)$ is unimodal. However the foregoing strategy comes with the drawback that it does not involve $\alpha$, so it removes the elegant possibility that the state-dependent model encompasses Poisson and power-law glitch activity in a single framework just by varying $\alpha$.

An alternative approach is to consider a family of functions, $\lambda(X; a)$ which tend to $\delta(X - 1)$ as $a \rightarrow 0$, e.g.
\begin{equation}
\label{eq:delfam}
{\lambda[X(t); a] = \frac{1}{a} \exp \left[\frac{X(t) - 1}{a}\right]}\ ,
\end{equation}
where $a$ is a constant, analogous to the dimensionless control parameter $\alpha$. Using \eqref{eq:delfam}, we find $\textit{p}(\Delta t) = \textit{p}(\Delta X) = \eta(\Delta X \mid \xp)$ for values of $a \lesssim 0.05$, as expected. However, for high values of $a$, $\textit{p}(\Delta t)$ becomes uniform if $\eta(\Delta X \mid \xp)$ is unimodal, as in \eqref{eq:gausseta}. Thus, like \eqref{eq:xp1}, this rate function does not produce both exponential and quasiperiodic waiting times easily by varying a single control parameter.

\section{Observational implications}
\label{sec:discussion}
\subsection{Waiting time and size PDFs}
A recent nonparametric analysis of the waiting time and size PDFs for the five most active glitchers confirms that there are two general classes of activity: exponential waiting times with monotonically decreasing glitch sizes (PSR J0534$+$2200 and PSR J1740$-$3015), and quasiperiodic (i.e. unimodal) waiting times with roughly Gaussian glitch sizes (PSR J0537$-$6910 and PSRJ0835$-$4510) \citep{Howitt2018}. One pulsar (PSR J1341$-$6220) falls somewhere between these two extremes. Due to the relatively small samples ($N \leq 42$), there is not much information in the nonparametric PDFs beyond this broad dichotomy. In particular we cannot say much about the exact functional form. For example, is a monotonically decreasing PDF truly a scale-invariant power law, or is it the tail of a broad PDF with a scale, e.g. a broad Gaussian? PSR J0534$+$2200 is a convincing power law over 2 dex, but other pulsars are less clear-cut.

Table \ref{tab:eta} is reductive; it does not encapsulate all of the information encoded in the long term statistics for all possible $\eta(\Delta X \mid \xp)$. Nonetheless it reveals some broad trends which can be compared to observed nonparametric PDFs. In particular, the Gaussian $\eta(\Delta X \mid \xp)$ produces exponential waiting times and monotonically decreasing glitch sizes for high $\alpha$, and unimodal waiting times and glitch sizes for low $\alpha$. If the message of the analysis by \citet{Howitt2018} is that size PDFs for the ``Poisson-like'' pulsars are strict power laws and not just broad, monotonically decreasing functions, the state-dependent Poisson process struggles to produce both power-law and unimodal glitch sizes just by changing the control parameter $\alpha$. On the other hand, if the samples are too small to be definitive about the exact functional form of $\textit{p}(\Delta X)$, then the Gaussian $\eta(\Delta X \mid \xp)$ does a fair job of capturing the two types of behaviour in the low-$\alpha$ and high-$\alpha$ regimes respectively.

\begin{figure}
	\centering
	\includegraphics[width=0.95\linewidth]{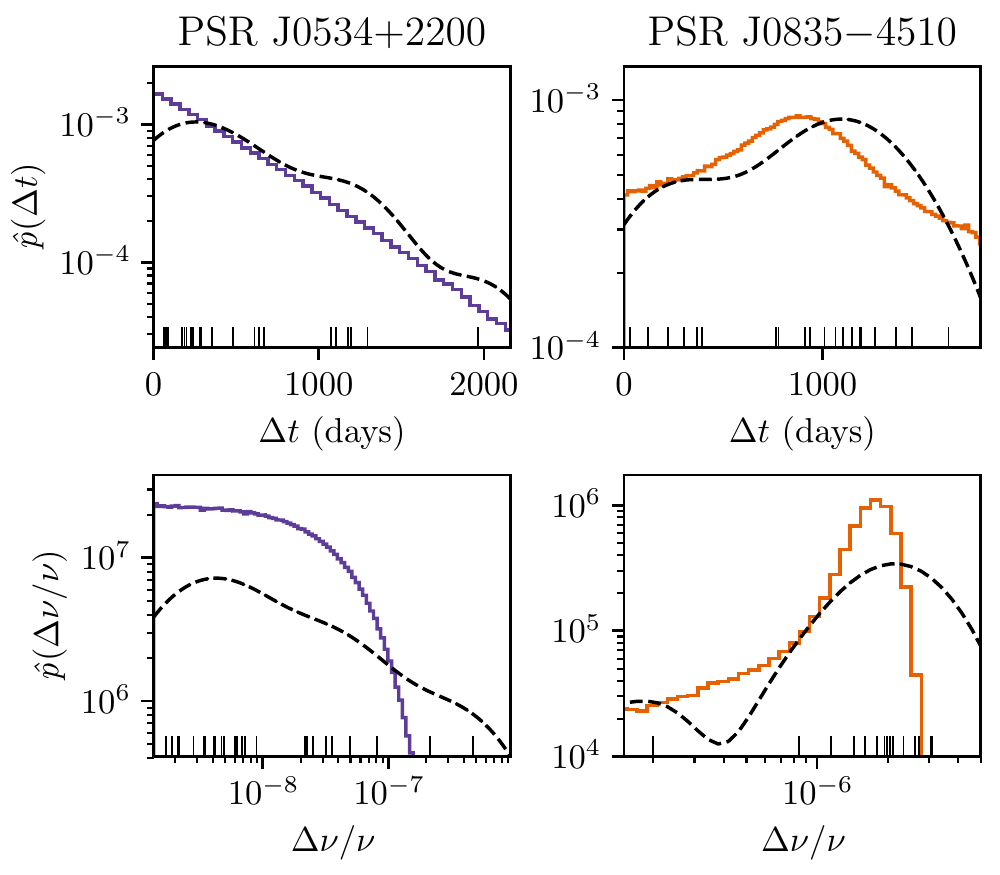}
	\caption{Nonparametric kernel density estimates (dashed black curves) of the waiting time (top row) and fractional glitch size (bottom row) PDFs for PSR J0534$+$2200 (left column) and PSR J0835$-$4510 (right column). Data points are denoted by thin black tick marks on the horizontal axes. Overlaid on the same plots are scaled PDFs generated using \eqref{eq:gausseta} and \eqref{eq:ndim_rate}, with $\alpha=15$ for the purple curves, and $\alpha=0.5$ for the orange curves. Parameters: $\mu_\textrm{G} = 0.5$, $\sigma_\textrm{G}=0.125$, $N=10^6$.}
	\label{fig:kdecomp}
\end{figure}

The aim of this paper is limited to exploring the behaviour of the theory under various input assumptions. We do not seek to fit the data for individual objects, partly because the samples available are small, and partly because the theory is idealized. However, to illustrate the points above, Figure \ref{fig:kdecomp} shows $\textit{p}(\Delta t)$ and $\textit{p}(\Delta \nu/\nu)$, where $\Delta \nu / \nu$ is the fractional glitch size, for PSR J0534$+$2200 (the canonical Poisson glitcher) and PSR J0835$-$4510 (the canonical quasiperiodic glitcher), compared against the theory. The measured PDFs are constructed using the nonparametric kernel density estimator \citep{Howitt2018}. The theoretical PDFs are generated using \eqref{eq:gausseta} and \eqref{eq:ndim_rate} and are scaled such that the mean of the generated waiting times (sizes) equals the mean observed waiting time (size). The $\alpha$ value for each object is selected as described in Appendix \ref{app:correl}. A Kolmogorov-Smirnov two-sample test indicates that the observed and generated waiting times are consistent with being drawn from the same distribution (p-values of 0.94 and 0.76 for PSR J0534$+$2200 and PSR J0835$-$4510 respectively). However the observed and generated sizes are not (p-values of $4\times10^{-6}$ and $7\times10^{-2}$). It would be premature to draw any conclusions from this illustrative exercise except to note that the shapes of the measured and theoretical PDFs are not entirely dissimilar. 

\subsection{Cross-correlations}
It is interesting to ask whether the size-waiting time cross-correlations can constrain the functional form of $\eta$, in the light of recent analysis of the observed correlation coefficients and their dependence on $\alpha$ \citep{Melatos2018}. We present a preliminary study of this issue in Appendix \ref{app:correl} while we await larger data sets. The main finding is that a significant backward cross-correlation between sizes and waiting times is expected for most unimodal functions $\eta$. No significant backward cross-correlations are observed to date. If this situation persists, as more data are collected, it would indicate either that we have $\alpha \lesssim 1$ in every pulsar, or that $\eta$ is not a unimodal function, in the context of the state-dependent Poisson process.

\subsection{Core and crust}
One goal of glitch modelling is to probe the relative roles played by the superfluid in the core and the crust of the neutron star. Recent calculations of entrainment between the neutron superfluid and nuclear lattice combined with a phenomenological two-fluid model imply that the angular momentum reservoir carried by the superfluid in the inner crust is not large enough to explain the ``regular'' glitches in pulsars  such as PSR J0537$-$6910, if the stress reservoir is completely emptied at each glitch \citep{Andersson2012, Chamel2012}. Two-fluid models have been applied to glitch trigger mechanisms \citep{Andersson2003, Glampedakis2009}, glitch rise times \citep{Sidery2010}, and measuring pulsar masses using glitch data \citep{Ho2015}.

In the state-dependent Poisson model applied to the superfluid vortex avalanches, the moment-of-inertia ratio of the crust and core, $I_\textrm{c}/I_\textrm{s}$, enters through the relation between $\Delta X^{(i)}$ and $\Delta \nu^{(i)}$, viz. equation \eqref{eq:delx_ratio}. As it is impossible to directly observe $\Delta X^{(i)}$, measuring $I_\textrm{c}/I_\textrm{s}$ is difficult. Maximum likelihood estimation of $I_\textrm{c}/I_\textrm{s}$ is possible in principle but it involves scanning over seven parameters at a minimum and falls outside the scope of this paper. The meta-model does not distinguish between superfluid in the inner crust and core; $I_\textrm{s}$ is proportional to the total angular momentum in both regions. Interestingly we find that the stress reservoir is not depleted totally under typical conditions in the state-dependent Poisson model (see figure 1 of \citealp{Fulgenzi2017}), i.e. we find $\Delta X^{(i)} \neq \xp$ regardless of the choice of $\eta$ or $\alpha$. In other words, the stress reservoir does not need to empty completely to produce quasiperiodic glitches.

\section{Conclusions}
\label{sec:concl}
Glitching pulsars broadly fall into two statistical classes: those with exponential waiting times and monotonically decreasing sizes, and those with unimodal waiting times and sizes. A microphysics-agnostic meta-model based on a state-dependent Poisson process can generate both quasiperiodic and exponential waiting times through varying the control parameter $\alpha$, when the conditional jump size distribution, $\eta(\Delta X \mid \xp)$, is unimodal. Likewise, the size PDFs generated by the model capture the broad features of the data. The size PDF is not a strict power law in the large-$\alpha$ regime, but it is unclear whether the relatively small glitch samples available to date absolutely require a power law either. This is a step forward from previous analyses with $\eta(\Delta X \mid \xp)$ of power-law form, which failed to produce unimodal $\textit{p}(\Delta t)$ and $\textit{p}(\Delta X)$ in the small-$\alpha$ regime.

Somewhat counterintuitively, altering the phenomenological rate function does not seem to be the best way to produce non-monotonic waiting times from the model. With the inputs tested in this paper the model struggles to generate both classes of glitching behaviour, unless the jump size distribution is changed from the power law used in previous work. 

Physically, if $\eta(\Delta X \mid \xp)$ is unimodal, there exists a characteristic size for the stress released at each glitch. There are ways to achieve this microphysically, of course, but matters are complicated by the spatially correlated nature of crustquakes and superfluid vortex avalanches, which leads naturally to scale invariance. Alternatively, one can interpret the results as evidence against one glitch mechanism governing all pulsars, or that the canonical view of glitches as a result of a marginally critical system is incomplete. 

\section{Acknowledgements}
\label{sec:ackn}
Parts of this research are supported by the Australian Research Council Centre
of Excellence for Gravitational Wave Discovery (OzGrav) (project number
CE170100004).



\newpage
\bibliographystyle{mnras}
\bibliography{sdpp_bib} 


\newpage
\appendix
\section{Size-waiting time cross-correlations}
\label{app:correl}

\begin{figure}
	\centering
	\includegraphics[width=0.95\linewidth]{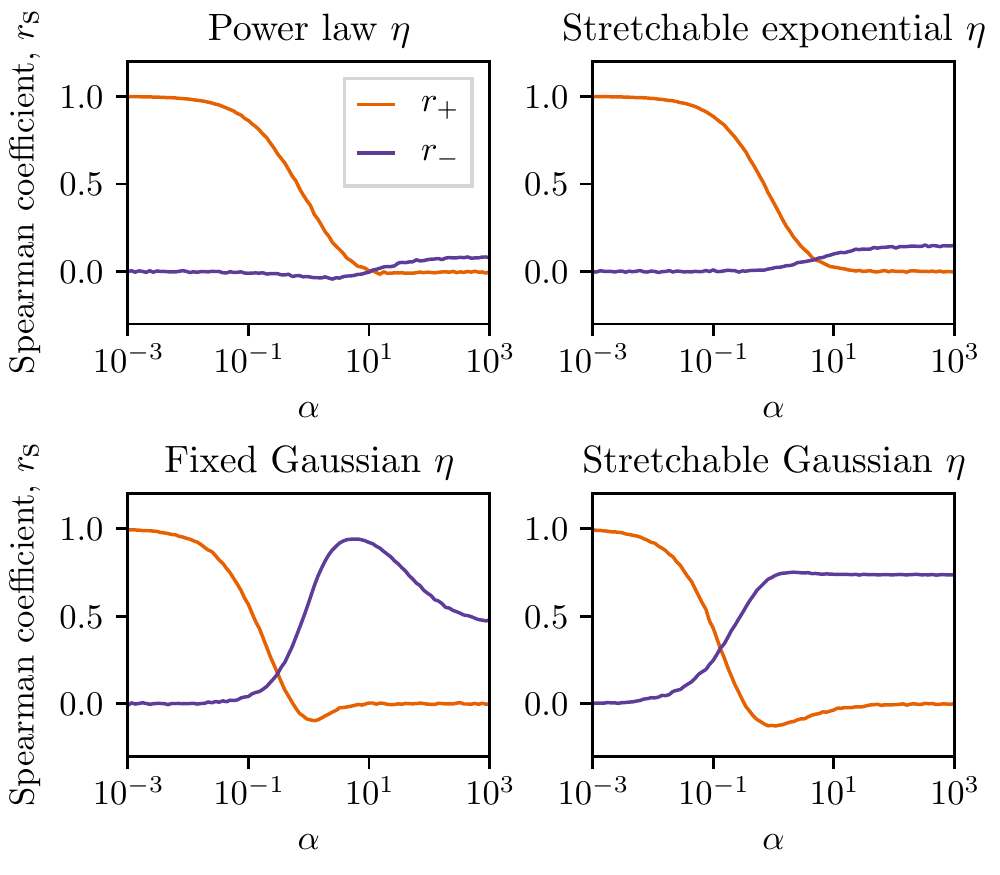}
	\caption{Forward ($r_+$, orange curves) and backward ($r_-$, purple curves) size--waiting-time cross-correlations \citep{Melatos2018} for different functional forms of $\eta$. Simulation parameters: 100 logarithmically spaced $\alpha$ values, $10^5$ glitches per $\alpha$ value,\; $\beta=10^{-2}$ for power-law $\eta$ (top-left panel), $\mu_\textrm{G}=0.5$ and $\sigma_\textrm{G}=0.125$ for Gaussian $\eta$ (bottom-left and bottom-right panels). A version of the top-left panel first appeared as Figure 13 in \citet{Fulgenzi2017}.}
	\label{fig:correls}
\end{figure}
The state-dependent Poisson process provides a framework to predict the correlations between the sizes and waiting times to (since) the next (previous) glitch, henceforth called the forward (backward) cross-correlation. In the fast spin-down regime ($\alpha \lesssim 1$), one obtains significant forward cross-correlations. The stress, $X(t)$, is driven rapidly to $\approx X_\textrm{cr}$ before each glitch. If the glitch is large, $X(t)$ takes longer to recover to $X(t) \approx X_\textrm{cr}$, than if the glitch is small. There is no significant backward cross-correlation in this regime, because $\eta(\Delta X \mid \xp \approx 1)$ is independent of the backward waiting time. On the other hand, in the slow spin-down regime ($\alpha \gtrsim 1$), cross-correlations show more varied behaviour, as they depend on the particular choice of $\eta(\Delta X \mid \xp)$.

Simulations bear out the trends described above, as we see in Figure \ref{fig:correls}. At low values of $\alpha$ there is a strong forward cross-correlation and no backward cross-correlation for monotonic and unimodal forms of $\eta(\Delta X \mid \xp)$. At high values of $\alpha$ neither the power law nor the stretchable exponential show significant forward or backward cross-correlations. Small glitches are favored for both these functional forms, rendering the restriction $\Delta X \leq \xp$ largely irrelevant. In contrast, a functional form for $\eta(\Delta \mid \xp)$ which favours larger glitches (e.g. the Gaussians in the bottom-left and bottom-right panels) produces significant backward cross-correlations for $\alpha \gtrsim 1$. In the slow spin-down regime, we have $X(t)\approx 0$, and the restriction $\Delta X \leq \xp$ plays an important role. More glitches return the reservoir close to zero, and a correlation between the size of the glitch and the time since the last glitch (i.e. the available reservoir) emerges. 
 
Figure \ref{fig:correls} indicates that we should expect to see some backward cross-correlations in pulsars with a unimodal $\eta$ but no forward cross-correlations in the regime $\alpha \gtrsim 1$. The fact that significant backward cross-correlations have not been observed so far may be a product of the small samples. Alternatively it may indicate that we have $\alpha \lesssim 1$ for all objects, or that $\eta$ is not unimodal. 

\begin{figure*}
	\centering
	\includegraphics[width=\linewidth]{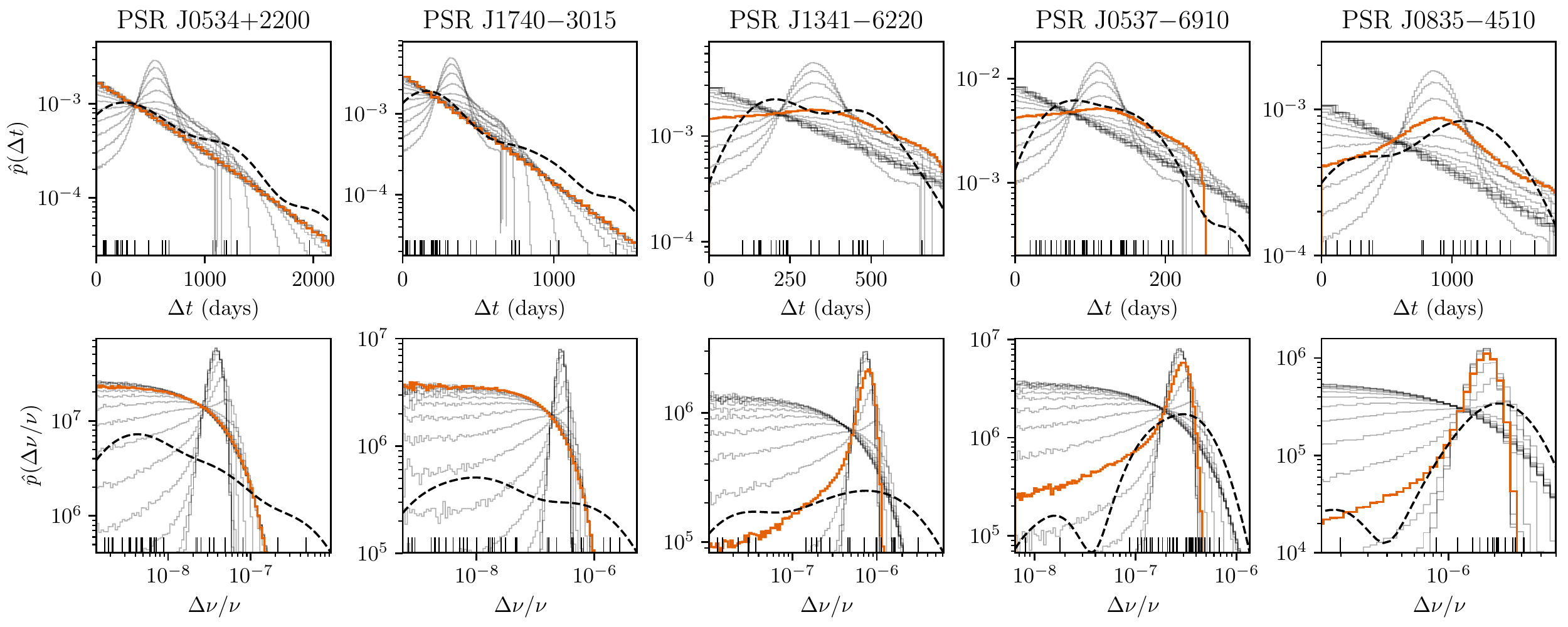}
	\caption{Nonparametric kernel density estimates (dashed black curves) of waiting time (top row) and size (bottom row) PDFs for the top five most active glitching pulsars compared with PDFs output by the state-dependent Poisson model. Black tick marks indicate measured data points. Overlaid on each panel are PDFs generated using \eqref{eq:gausseta} and \eqref{eq:ndim_rate} scaled such that the mean equals the mean of the observations. Each grey curve represents a simulation with one of 20 logarithmically spaced $\alpha$ values, ranging from $\alpha=0.01$ to $\alpha=100$. The orange curves correspond to the $\alpha$ value that minimises the Kolmogorov-Smirnov two-sample statistic. Simulation parameters: $N=10^6$, $\mu_\textrm{G}=0.5$, $\sigma_\textrm{G}=0.125$}
	\label{fig:all_kdecomp}
\end{figure*}

Precisely inferring parameters, such as $\alpha$, from available glitch data is computationally expensive and falls outside the scope of this work. It is also premature given the small samples available. However we can make some broad statements by matching the output waiting time and size PDFs of the state-dependent Poisson model to observed PDFs. Figure \ref{fig:all_kdecomp} shows the nonparametric kernel density estimates for the waiting time and size PDFs for the five pulsars with the most recorded glitches. Overlaid on each plot are PDFs generated with the state-dependent Poisson model, using the Gaussian $\eta$ defined in \eqref{eq:gausseta} and the standard rate function defined in \eqref{eq:ndim_rate}, each with a different value of $\alpha$. PDFs are generated for 20 logarithmically spaced $\alpha$ values, ranging from $\alpha=0.01$ to $\alpha=100$. Each set of $N=10^6$ glitches is scaled such that the mean waiting time equals that of the observed glitches (and equivalently for sizes). The generated PDF highlighted in orange is the one that has the smallest Kolmogorov-Smirnov two-sample statistic for the waiting time distribution. Of the values tested, the statistic is minimised for $\alpha=15$, $\alpha=100$, $\alpha=0.8$, $\alpha=0.8$ and $\alpha=0.5$ for PSR J0534$+$2200, PSR J1740$-$3015, PSR J1341$-$6220, PSR J0537$-$6910 and PSR J0835$-$4510 respectively. 

\bsp	
\label{lastpage}
\end{document}